\documentclass[9pt, twocolumn]{extarticle}

\usepackage{authblk}
\usepackage{amssymb}

\makeatletter
\renewcommand{\maketitle}{\bgroup\setlength{\parindent}{0pt}
\begin{flushleft}
  \textbf{\@title}

  \@author
\end{flushleft}\egroup
}
\makeatother

\title{\Huge 
Towards fair roads \textendash \; Why we should \& how to improve the fairness in traffic engineering
\vspace{0.5cm}}
\author{
\Large 
 Kevin Riehl$^{1,*}$ ,
 Anastasios Kouvelas$^{1}$ , 
 Michail A. Makridis$^{1}$ \\
\tiny 
[*] Corresponding author: kriehl@ethz.ch \\
[1] Traffic Engineering Group, Institute for Transport Planning and Systems, ETH Zurich, Stefano-Franscini-Platz 5, Zurich, 8093, Switzerland
}

\date{\today}

\usepackage[showframe=false, margin=.65in]{geometry}

\usepackage{abstract}
\renewcommand{\abstractname}{}    

\usepackage{lipsum}

\usepackage{afterpage}

\renewenvironment{abstract}
 {\small
  \begin{center}
  \bfseries \abstractname\vspace{-.5em}\vspace{0pt}
  \end{center}
  \list{}{%
    \setlength{\leftmargin}{10mm}
    \setlength{\rightmargin}{\leftmargin}%
  }%
  \item\relax}
 {\endlist}

\usepackage[utf8]{inputenc}

\usepackage{wrapfig}
\usepackage[lofdepth,lotdepth]{subfig}
\usepackage{booktabs}

\usepackage{rotating}
\usepackage{caption}

\usepackage{csquotes}

\usepackage{todonotes}

\usepackage{stfloats}

\usepackage[T1]{fontenc}
\usepackage{textcomp}
\usepackage{times}

\usepackage{framed} 

\usepackage[citestyle=nature, 
            maxcitenames=2,
			bibstyle=nature, 
			language=auto,
			url=false,
			backend=bibtex,
			doi=false,
			isbn=false]{biblatex}
	\renewbibmacro*{volume+number+eid}{%
  \printfield{volume}%
  \setunit*{\addnbspace}
  \printfield{number}%
  \setunit{\addcomma\space}%
  \printfield{eid}}
\DeclareFieldFormat[article]{number}{\mkbibparens{#1}}
\usepackage{setspace}
\usepackage{color}
\definecolor{black}{gray}{0} 

\usepackage[colorlinks=true,linkcolor=black,citecolor=black]{hyperref}

\usepackage[]{fancyhdr}
{\renewcommand{\markboth}[2]{}
\fancyhead[R,RO]{}
\fancyhead[L,LO]{}
\pagestyle{fancy}
\fancyhead[RO,R]{
\textit{
\\\hrulefill\\
Towards fair roads \textendash \; (Riehl et al. 2024) \; \textendash \; A Submission To: npj | sustainable mobility and transport (2024) }
}

\usepackage{algorithm}
\usepackage{algpseudocode}
\newcounter{algsubstate}

\usepackage{tabularx}
\newcolumntype{s}{>{\hsize=.5\hsize}X}

\usepackage{ntheorem}

\begin{document}

\twocolumn[
    \begin{@twocolumnfalse}
        \maketitle
        \begin{abstract}
            Traffic engineering aims to control infrastructure and population behavior to achieve optimal usage of road networks. Fairness is fundamental to stimulate cooperation in large populations, and plays an important role in traffic engineering, as it increases the well-being of users, improves driving safety by rule-adherence, and overcomes public resistance at legislative implementation.
            Despite the importance of fairness, only a few works have translated fairness into the transportation domain, with a focus on transportation planning rather than traffic engineering.
            This work highlights the importance of fairness when solving conflicts of large populations for scare, public good, road-network resources with traffic engineering, and establishes a connection to the modern fairness theories.
            Moreover, this work presents a fairness framework that serves when designing traffic engineering solutions, when convincing in public debates with a useful, argumentative tool-set to confront equity considerations, and enables systematic research and design of control systems.
            \hfill \break \;
            \newline
            \newline
            \newline

        \end{abstract}
    \end{@twocolumnfalse}
]



\section{Introduction}
The usage of transportation infrastructure is vital for a meaningful, joyful, fulfilling, and dignified life in today's society, as it enables access to the job market, educational \& recreational activities, social participation, and marketplaces~\cite{karner2023emerging}.
For many people worldwide, road transportation is the only accessible and affordable means of transportation.
The access to the road infrastructure is shared by many users, 
and the utility of the infrastructure depends on its usage; when being used over capacity, congestion arises and the utility diminishes.
Ever growing population and demand, and the induced over-consumption of shared transportation infrastructure is a growing issue in mega cities worldwide, leading to conflicts between the users, such as congestion~\cite{verhoef2010economics}.

In general, two strategies exist to resolve these conflicts: transportation planning and traffic engineering.
Transportation planning is concerned with designing future infrastructure to meet growing demand patterns, while traffic engineering is concerned with controlling \& actuating the existing infrastructure as efficiently as possible.
Increasing the supply by building more infrastructure can only offer a temporary solution, as it triggers yet another increase in demand, in the long run.
Therefore, traffic engineering must find better ways to resolve the conflicts arising from existing, limited infrastructure~\cite{roess2004traffic}.
Contrary to transportation planning, traffic engineering must react to given situations in real time, plans automatically, and affects users in the short-term rather than the long-term.
The aim of traffic engineering is the management of road transportation systems in order to achieve congestion free, efficient traffic, optimal utilization of infrastructure (at capacity), and conflict management.
Infrastructural control (traffic operations, e.g. with traffic lights) aims to manage traffic in a way to avoid or mitigate congestion.
Behavioral control (demand management, e.g. with congestion pricing) affects the behavior and decision making process of users using economic instruments, and aims to keep usage at sustainable levels reflecting infrastructural capacity limitations.

Traffic engineering is challenging, as it requires the alignment of system and user goals, which can be conflicting.
The introduction of traffic engineering solutions, can only be achieved via policy-making, which requires public acceptance.
One of the major, public concerns when debating on the introduction of solutions are equity considerations and social feasibility~\cite{provoost2023design}, where equity considerations usually outweigh potential efficiency improvements~\cite{Varian2018}.
For example, even though ramp metering~\cite{yin2004note,li2016efficiency}, perimeter control~\cite{aboudolas2013perimeter}, and congestion pricing~\cite{eliasson2016congestion} provide many potential benefits for the society, they have been heavily criticised in controversial debates of the public which hindered their widespread use in many cases.

Over the past decades, the vast majority of intellectual energy, societal debates, and public resources have been dedicated to solving congestion and improving transportation efficiency purely from a system perspective~\cite{de2011handbook,goodwin1997solving}. 
Higher efficiency is beneficial in general, as it avoids, mitigates or reduces congestion, and improves the average passengers travel time. 
However, this mitigation activity will not only generate beneficiaries amongst the users.
A purely efficiency-driven approach can be highly problematic, as misspecified designs of traffic engineering solutions harbor the threat of creating systematic inequalities between the users, unacceptably long waiting times, or excessive queue forming.
Besides, transportation efficiency is a multidimensional concept, and optimizing for one goal (e.g. reducing delays), can deteriorate conflicting goals (e.g. more pollution, more energy consumption).
As the transportation justice movement states, transportation inequalities affect access to job opportunities, education, and other social activities, and thus reinforce inequality patterns over the course of time~\cite{martens2012justice,inwood2015we}.  

The significance and importance of fairness was largely overlooked by scholars and practitioners alike in the context of traffic engineering.
Fairness heavily determines the success of traffic engineering solutions.
Solutions that are perceived as fair, are more likely to be approved by political processes and adapted by the population~\cite{Varian2018}.
Fairness is crucial for safety; without the consideration of fairness, users that perceive unfairness (e.g. very long waiting times at traffic lights) are observed to disrespect rules \& structures and circumventing systems, leading to security incidents~\cite{hendriks2021equity,wissema2002driving,huntley1983so,jusoh2017distributed}.
Fairness is fundamental to stimulate cooperation in populations with many individuals, and to generate acceptance for social structures and rule-adherence behavior~\cite{gurney2021equity}.
Perceived fairness affects the psychological and physical well-being of users as a hygiene factor~\cite{gurney2021equity,jackson2006linking}.

Even though fairness has been extensively examined in other societal-relevant disciplines such as public healthcare, housing, environmental and social sustainability, only few works have translated fairness into the transportation domain, and most of those focus on transportation planning rather than traffic engineering~\cite{pereira2017distributive}. 
The few studies in traffic engineering are limited by the unsystematic study of fairness, by the attachment to specific ideologies of fairness (Egalitarianism), by the employment of over-simplifying measures, by the lack of connection with the fairness literature, and by a focus on the vehicles / drivers only.
Therefore, it is the aim of this work to establish a link between modern fairness theories and algorithmic control, and to empower systematic research on fairness in the domain of traffic engineering.

We propose a quantitative, ideology-free, distributive, and pragmatic fairness framework that shall be useful when designing, discussing and implementing infrastructural and behavioral control systems.
Traffic engineering solutions are modelled as resource allocation mechanisms, and identify delays, pollution, financial costs, and budgets as fairness-relevant resources in the spirit of modern fairness theories like Rawls~\cite{rawls1971atheory}, Walzer~\cite{walzer1983} and the capabilities approach~\cite{nussbaum2011creating,sen2008idea}.
Following the Nicomachean transactional view~\cite{wolf2002aristoteles}, we discuss distributive fairness in a quantitative way, and propose fairness measures for diorthotic fairness (welfare functions), and for dianemetic fairness (concentration \& dispersion metrics).
Moreover, two case studies demonstrate the usefulness and feasibility of this approach.
The results of the case studies show, that both forms of control improve the fairness levels overall, and that fairness and efficiency do not exclude each other. 
However, Egalitarian ideology is the most conflicting with efficiency, and at the same time the most widely used measure in the literature, which indicates the need for a more differentiated discussion on fairness.
Finally, a dendrogram analysis of various proposed fairness measures reveals clusters that correspond to the underlying fairness ideologies of our proposed framework.

The remainder of this work is as follows.
Section~\ref{sec:lit-review} reviews the literature on fairness, fairness in transportation in general, and fairness in traffic engineering in particular.
Section~\ref{sec:framework} develops the distributive fairness framework for road traffic engineering.
Section~\ref{sec:casestudies} demonstrates the usefulness of the developed framework at two examples: signalized intersection management, and static road pricing.
Section~\ref{sec:conclusions} concludes this work with a summary.


\section{Literature Review}
\label{sec:lit-review}

Discussing fairness is challenging, as there is no consensus on a clear definition, and defining fairness heavily depends on social and cultural contexts.
In the context of traffic engineering, previous work on fairness is limited, in that the previous studies: (i) mostly discuss fairness in a qualitative rather than quantitative way, (ii) are limited by Egalitarian ideology, (iii) are not connected to fairness literature, (iv) solely employ crude, oversimplifying metrics that measure the distribution's dispersion (mostly standard deviation), and (v) focus on the situation of vehicles / drivers only, overlooking the negative externalities of traffic on residents or other stakeholders.
Hence, there is the necessity for more systematic, quantitative research on fairness, and a dedicated framework to connect traffic engineering and the fairness literature is necessary to involve considerations from multiple ideologies' point of view.

In the first part of this section, we motivate fairness as a requirement for cooperation \& as an evolved trait of social animals, discuss the backgrounds of this term, and demarcate the terminology against equity, equality, and justice.
Afterwards, we present the most recognized fairness theories from the modern, political-philosophy literature, elaborate on the activist, social justice movement, and close with an overview on the fairness of markets.
The second part reviews previous work on fairness in the general transportation domain, discourses political, social justice movements related to (public) road transportation, and elaborates on one of the most recognized fairness theories in the transportation domain, which is concerned with transportation planning: "transportation justice".
The third part systematically reviews and compares the few, previous works on infrastructural and behavioral control in traffic engineering, and identifies the limitations of these works.

\subsection{Fairness}

\subsubsection{General introduction to fairness}
A discussion of fairness is challenging in general, as fairness is an intangible, philosophical idea, that must be discussed specific to social and cultural contexts. 
More than two millenniums of philosophical discussions across different human civilizations have not yet resulted in consensus~\cite{goppel2016handbuch}.
What's more, the many theories on fairness in the philosophic \& political-philosophic discussion are rather transcendental, transitive-comparative, and qualitative in nature, which is of little help for traffic engineering purposes. 

From an evolutionary biology point of view, sensitivity to fairness is a behavioral trait that evolved in social animals, whose survival depends upon cooperation, and a social group.
The perception of fairness can be observed for many social animals, such as vampire bats~\cite{carter2015social}, capuchin monkeys~\cite{brosnan2014evolution}, and humans~\cite{corradi2016cross}.
On the individual level, the subjective perception of fairness is found to be important for the psychological well-being~\cite{gurney2021equity} and physical health~\cite{jackson2006linking}.
Social brains react to the lack of fairness with a feeling of disgust, triggered by the insula~\cite{corradi2016cross},
where the reaction is stronger for a subjectively perceived, unfair disadvantagement compared to an unfair advantagement.
Moreover, perceived fairness affects how individuals engage relationships and interact with each other~\cite{de2010procedural}.
On a population level, fairness is fundamental to building strong communities amongst individuals, to encourage cooperation and social rule-following~\cite{gurney2021equity}.
Shared, tribe-specific norms and culture determine the perception of fairness~\cite{hitti2011social}.
The perception of fairness is not only happening on the individual level, but it is described also as an emergent, cognitive phenomenon on a population level~\cite{kozlowski2000multilevel}.
What's more, fairness strengthens social cohesion, reinforces identification and trust with the group, reduces conflicts, and yields higher (economic) outcomes~\cite{de2010procedural}.  

Fairness is considered as the foundation for human cohabitation, and therefore widely discussed in philosophy, ethics, sociology, politology, economics, and religion science, as a multidisciplinary term.
Hume~\cite{hume1739} requires three conditions to discuss the question of fairness: (i) scarcity of resources, (ii) a conflict of interest, and (iii) a relative balance of power between the negotiating parties.
Goppel et al.\cite{goppel2016handbuch} differentiate types of fairness theories, that vary in their level (individual vs. institution vs. nation vs. global), scope (temporal, spatial, recoccurence), focus (procedure, distribution, transactional, result, need, criminal justice), perspective (ex-ante, ex-post), and ideologies (outlined in Supplementary Table 3). 
The meaning of fairness changed over time; in the ancient philosophical discussions it was considered as an individual's virtue and one of the most important traits of an iconic deity; beginning from the renaissance the deity was replaced with a form of natural and universalistic law; during the period of enlightenment, philosophical discourse distanced from religion science and started to focus on a ratio based reasoning; modern discussions on fairness are of political and economic nature and focus not only on individuals but on institutions and processes (such as markets)~\cite{johnston2011brief}.
For an extensive overview on fairness the reader is recommended to review~\cite{goppel2016handbuch}.

Fairness is a hypernym for equality, that refers to an allocation where each individual is given the exact same amount of resources (Egalitarian fairness ideology), neglecting contribution, need, preferences or utility.
The terms of fairness and justice are used interchangeably in the literature~\cite{goppel2016handbuch}. 
Equity is a terminology for fairness that is predominantly used in economic considerations.
Justice is a terminology for fairness that appears in political debates, such as the social justice and environmental justice movements.

Colquitt's fourfold fairness model~\cite{colquitt2001dimensionality} focuses on describing the fairness perception process and distinguishes distributive, procedural, interpersonal and informational fairness. Moreover, fairness climate models for the fairness perception on a population level are increasingly studied~\cite{li2009fairness,rupp2007justice}.

\subsubsection{Transactional framework of distributive fairness}
At its core, fairness (and especially distributive fairness) is a question of allocation (distribution) of resources across a population as a result of transactions.
Resources do not only include desirable, value-generating resources, but can also refer to undesirable resources with negative value, to burdens, or to penalties.

\begin{figure} 
    \centering
    \includegraphics[width=1.0\linewidth]{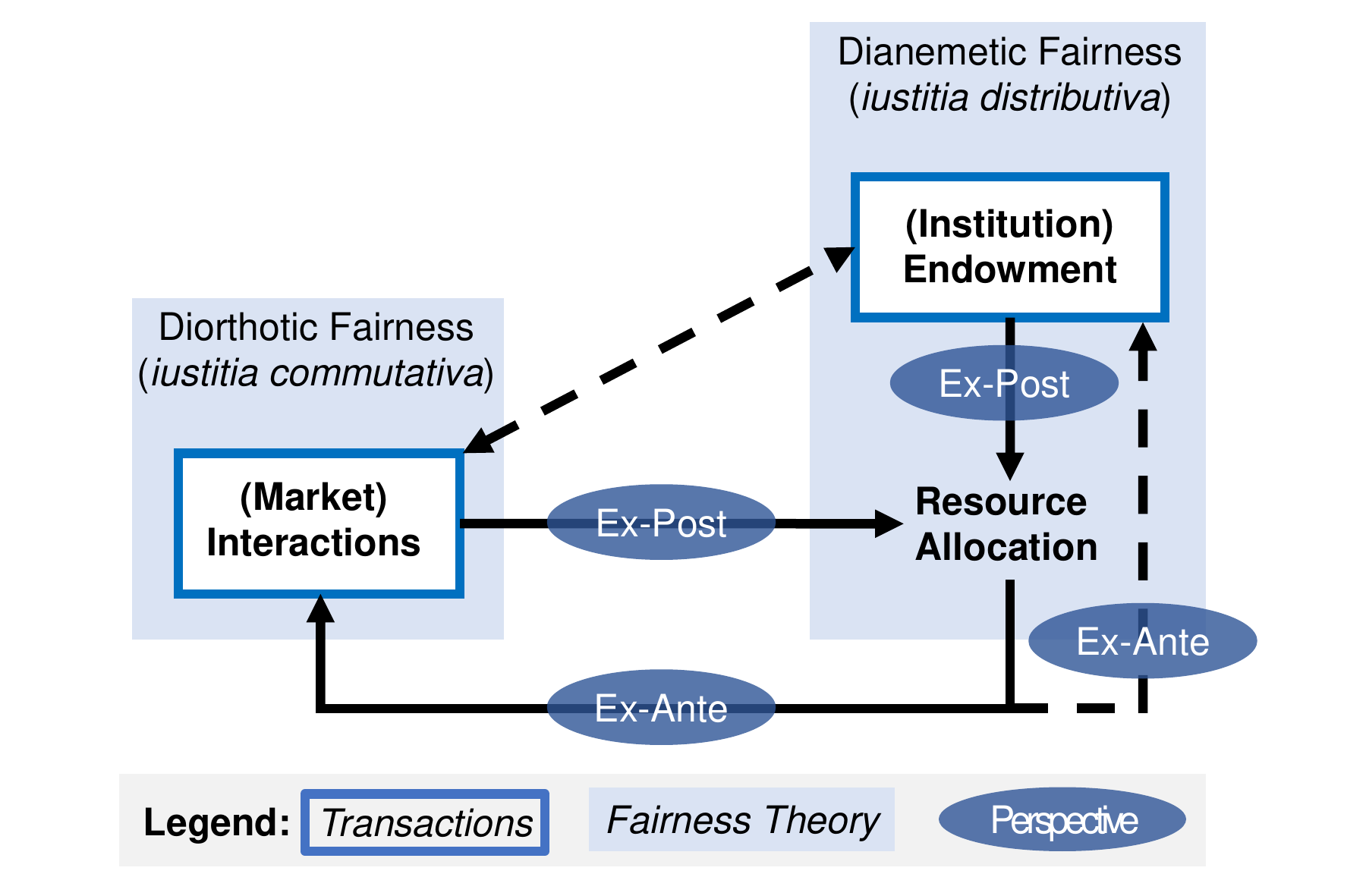}
    \caption{
        \textbf{Transactional fairness framework}. \\
        The Aristotelian, Nicomachean, transactional fairness theory discusses fairness in the context of resource transactions, and differentiates dianemetic and diorthotic fairness. 
        Diorthotic fairness refers to transactions as interactions between individuals, e.g. on markets. Dianemetic fairness refers to transactions from an authority to a population, e.g. governmental resource distribution.
    }
    \label{fig:fairness_process}
\end{figure}

Following the Aristotelian, Nicomachean, transactional fairness theory, fairness discussions can be grouped into dianemetic and diorthotic fairness, as shown in Fig.~\ref{fig:fairness_process}. 
Transactions can be interactions between different types of parties under different circumstances; this can include trades on markets, endowment of resources from government to population, participation in governmental decision making processes, or a court tribunal.
Dianemetic, distributive fairness is concerned with transactions in which a population is endowed with resources from an authority. Usually this discussion is concerned with governmental institutions that endow a population, for example in the case of subsidies or societal redistribution processes. The initial situation can also be considered as an endowment.
Diorthotic, commutative fairness is concerned with transactions between individuals (or groups) of a population. This usually includes trades on markets.
Nozick~\cite{nozick1974anarchy} bridges the gap between dianemetic and diorthotic fairness, as he argues that fair, diorthotic transactions can only be fair, if the allocation in the initial situation (dianemetic) has been fair already.
The fairness perspectives ex-post and ex-ante describe what matters for the fairness of a resource allocation; the first one focuses only on the output of a transaction, the second one focuses only on the input of a transaction, and of course a combination of both is possible.
Governments and markets stand in a relational interplay, as markets activity determines the economic power of a government, and governments can use their power to control markets (e.g. with taxation and redistribution). 

\subsubsection{Modern fairness theories and social justice}

Dworkin's equality of resources~\cite{dworkin1981equality,dworkknork} is an example for a Luck-egalitarian view on distributive justice~\cite{knight2013luck}.
Similar to the Liberal and Libertarian point of view, markets are considered as fair, property rights and freedom are cherished concepts, and "as few government as necessary" is a guiding principle.
Luck-Egalitarianism acknowledges that persons' skills and abilities are not equal, and thus advocates governmental intervention to compensate for undeserved misfortune that impacts their interests.
The outcome of societal, random, resource-allocating processes (such as markets) is considered fair, as long as every participant had similar initial conditions. Disparaties are advocated, as long as decisions of free individuals led to those.

Walzer's spheres of justice~\cite{walzer1983} is a Communitarist, distributive justice theory.
Communitarianism refuses any universal principle, and demands context-specific principles.
Walzer advocates different principles for goods of different levels of social meaning in different spheres of justice (complex equality).

One of the most recognized, modern theories of fairness is John Rawls' contractual, Egalitarian-Liberal work~\cite{rawls1971atheory}, that advocates disparities, as long as they enable the best possible situation for the least-advantaged (difference principle). 
Rawls argues, that fair allocations result from fair contracts, where fair contracts are contracts accepted and agreed upon by a population of individuals in an "original position" of equality behind the "veil of ignorance". 
In the original position, individuals do not know which social status, economic power and personal capabilities they will have in the real world (a market or society). As a consequence, as Rawls argues, individuals in the original position will consider achieving the best outcome for the worst case scenario (situation of the least advantaged) when designing contracts. Harsanyi~\cite{harsanyi1975can} however argues, individuals will consider the best average scenario.

Nozick's "anarchy, state and utopia"~\cite{nozick1974anarchy} is a Liberal answer to Rawlsian theory.
Nozick advocates a minimal state, that is \textit{"limited to the narrow functions of protection against force, theft, fraud, enforcement of contracts"}~\cite{maloberti2010squirrel}; besides that, the freedom of markets are defended against governmental intervention, and any form of profit taxation is criticized~\cite{nozick1974anarchy}.

The capabilities approach by Sen~\cite{sen2008idea} and Nussbaum~\cite{nussbaum2011creating} was developed as an alternative approach to welfare economics.
Rather than focusing on the right or freedom to achieve self-determined, dignified lives worth living, the capabilities approach focuses on the actual capabilities to achieve such a life.
The approach distinguishes functionings and capabilities. 
Functionings are desirable achievements, including wealth, health, nourishment, education, safety etc.  
Humans are equipped with goods, translate them based on individual-specific factors into a capability set, and then decide on how to use these capability sets to achieve functionings.  Thus, capabilities can be seen as potential functionings that individuals can reach.
For instance, a physically-impaired person needs more resources than a healthy person in order to achieve the same functioning (e.g. a travel). 

Social justice is an activist, political grass-root movement that originates from the empirical observation, that there are significant and systematic inequalities between certain groups of the society, for example racial segregation and discrimination, gentrification, sexism, ableism or ageism~\cite{inwood2015we,hananel2016justice}. Political activist groups demonstrate for systematic redistribution to achieve social mobility and emancipation from discriminative structures. These movements have in common, that they are system-critic, and argue that systematic discrimination reinforces and reproduces inequalities. 

\subsubsection{The fairness of markets}

The fairness of markets was closely linked with the philosophical, political and economic discussions of fairness.
Aristotle argues, a diorthotic transaction is fair, when the exchanged resources are of equal value, where value is not specified into more detail (e.g. value for the user, value for the trader, ...).
Albertus Magnus and Thomas Aquinas~\cite{chenu1954introduction} introduced the term of a fair price for transactions at monetary markets. The fair price primarily reflects the efforts for the generation of the resource, but can also include marginal profits of traders.
The school of Salamanca~\cite{liebrand2023francisco} puts the term of a fair price equal to the market price, assuming efficient, ideal markets.
Adam Smith's theory of the invisible hand~\cite{smith1937wealth} claims, that any selfish, egoistic behavior and any price in transactions is fair, as free markets lead to societal optima as a result. 

Besides market failure and arbitrage, purposeful phenomena such as price differentiation, dynamic pricing, price discrimination, and personalized pricing were heavily discussed in the literature. The fairness of markets is therefore closely linked with the fairness of prices~\cite{richards2016personalized, lee2011perceived, maxwell2002rule}.

\subsection{Fairness in Transportation}
Transportation shifted from a luxury into a necessity in industrialized societies~\cite{martens2016transport}. 
Access to transportation resources is vital for a meaningful, joyful, fulfilling, and dignified life, as it enables access to the job market, recreational activities, social participation, and marketplaces~\cite{karner2023emerging}.
Transportation justice is the consideration of fairness in questions of transportation, and evolved from the fields of social justice and the civil rights movement~\cite{sanchez2008assessing,hananel2016justice}.
Political, activist movements in California (transportation justice movement of Los Angeles), Atlanta (pray for transit), and Flanders in Belgium (basic accessibility debate) were recognized and inspired by the political-philosophy literature of the recent~\cite{vanoutrive2019just,karner2019pray}.

The discussions on fairness in transportation can be grouped into descriptive works that report disparities and the disadvantagement of specific groups (by race~\cite{karner2023emerging,johansen2021decolonizing,inwood2015we}, economic power~\cite{blumenberg2017social,karner2019pray,karner2020transportation}), and prescriptive, reformative works that advocate new forms of transportation planning~\cite{martens2016transport,vanoutrive2019just}. Besides, works elaborate on land-use~\cite{karner2016planning,martens2012justice,shoup2021high}, emissions, health \& environment~\cite{hansmann2023transportation,forkenbrock1999environmental}. A central concept that appears in most works are accessibility and potential mobility; moreover, most works share that they focus on the public transport.

Karel Martens' pioneering work "Transport justice: Designing fair transportation systems"~\cite{martens2016transport} is one of the most recognized fairness theories in the transportation domain. Martens bridges the gap between political-philosophy theories on fairness (Walzer's spheres of justice, Rawls' difference principle, and Sen\&Nussbaum's functional capabilities approach) and transportation. His work contributes threefold: (i) it criticizes traditional cost benefit analysis, (ii) it proposes a new form of transportation planning that emphasizes a strong involvement of disadvantaged groups into the planning process, and (iii) it advocates the concept of accessibility and Sufficientarianism.
Traditional cost benefit analysis follows an utilitarian approach aiming to maximize the benefits of the greater good (e.g. by considering minimal waste of value of time). Martens argues that this approach has a systematic bias, as it neglects that demand choices follow supply patterns and hence reinforces economic inequalities by favoring high income minorities in an undemocratic manner. Following Walzer, Martens identifies the social meaning of transportation goods in two concepts. Potential mobility refers to the ease with which a person can move through space. Accessibility refers to the potential of opportunities for interactions. Connecting these concepts with Rawlsian, Dworkian and Capabilities approach, he argues that the assessment of fairness shall be first and foremost a measurement of accessibility. Finally, he proposes a Sufficientarian theory of fairness for transportation, where a critical minimum of accessibility (the sufficiency threshold) must be guaranteed for each member of the society. However, he leaves the definition of a sufficient threshold open, and argues it must be determined and discussed in a democratic process.

While most initial works were rather qualitative, transcendental and comparative in nature, proposed quantitative measures for transportation planning become increasingly sophisticated in the transportation justice literature~\cite{karner2023emerging}.
The accessibility concept has been quantified in different measures, such as the potential mobility index~\cite{martens2012justice,martens2016transport}, the logsum-based access measure~\cite{bills2022towards}, a measure that captures supply and demand of shared mobility~\cite{desjardins2022examining}, a measure for home delivery services~\cite{figliozzi2021home}, and measures in the context of the 20-minute city~\cite{calafiore202220}.
Moreover, the maximum necessary travel time, distance, and expense to reach relevant locations for all residents of a specific area was discussed in~\cite{hananel2016justice}.

\subsection{Fair Traffic Engineering}

\subsubsection{Fair infrastructural control}
A comparative summary of the few works from traffic engineering that explicitly approach fairness in infrastructural control can be found in Table~\ref{tab:traffic_engineering_works}.
Most works identify the distribution of delays and queue lengths as relevant to fairness, which implies distributive fairness considerations~\cite{kesten2013analysis}.
Besides, there is an ongoing debate on the relationship between efficiency and equity.
Some works argue there is a goal conflict, and thus a trade-off between efficiency and equity must be drawn~\cite{levinson2004measuring,amini2016new,jusoh2017multi,kotsialos2001optimal,zhang2004optimal}, while others do not find that efficiency and fairness necessarily play a zero-sum-game ~\cite{kesten2013analysis,li2016efficiency,zhang2010access}.

In addition to that, several works draw parallels between computer science \& network engineering, and employ metrics from these domains, most prominently Jain's equity metric~\cite{hendriks2021equity,amini2016new,jain1991art}.
What's more, \cite{levinson2004measuring,hendriks2021equity} emphasize the importance of perceived rather than factual delays, and advocate forms of weighted delay optimization where shorter delays are considered under-proportionally important when compared with longer delays.

In the context of signalized intersection management, the distribution of delays is central to fairness considerations and usually fairness is understood as minimizing the standard deviation of delays~\cite{hendriks2021equity,yan2020efficiency,raeis2021deep,shirasaka2023distributed}.
\cite{raeis2021deep} distinguishes time delays of vehicles, and green times of movement phases.
Moreover, the distribution of queue lengths is also taken into account~\cite{li2012equity,huang2022fairness}, and fairness is understood as adherence to a scoping range for the queue lengths.
\cite{hendriks2021equity} contrasts a variety of equity measures and links how aspects of traffic light control algorithms correspond to certain theories of fairness.

In the context of ramp metering, \cite{levinson2006ramp} differentiates temporal (different freeway entrance times) and spatial (different on-ramps) inequity of delays.
\cite{levinson2004measuring} does not explicitly measure fairness in order to integrate fairness into optimization, rather they apply a non-linear function to weight delays, where longer delays have a stronger weight than shorter delays.
\cite{meng2010pareto} considers a spatial equity index, where delays of different ramps rather than different vehicles are considered.
Dispersion of delays is used to assess fairness in~\cite{li2016efficiency,kesten2013analysis,lu2017coordinated}.
\cite{amini2016new} considers a Nash bargaining solution and ~\cite{kotsialos2001efficiency} investigates spatial variation.

In the context of perimeter control, distribution of delays and queue lengths are considered as relevant to fairness. 
Alpha-fair~\cite{moshahedi2023alpha} and \cite{aboudolas2013perimeter} balance queue length and delay distribution as part of their perimeter control optimization and emphasize the importance of equity.

\begin{table*}
    \centering
    \begin{tabular}{l|m{5.0cm}|m{5.5cm}}
        \textbf{Control} & \textbf{Distribution of what} & \textbf{Fairness Metric} \\
        \hline
        
        & & \\
        Signalized intersection management & 
        \begin{itemize}
            \item Vehicle delay~\cite{hendriks2021equity,yan2020efficiency,shirasaka2023distributed} 
            \item Movement phase delay~\cite{raeis2021deep} 
            \item Queue length~\cite{li2012equity,huang2022fairness} 
        \end{itemize} & 
        \begin{itemize}
            \item Standard deviation ~\cite{hendriks2021equity,yan2020efficiency,raeis2021deep,shirasaka2023distributed} 
            \item Range~\cite{raeis2021deep} 
        \end{itemize} \\
        
        & & \\
        Ramp metering \& (Variable) speed limits & 
        \begin{itemize}
            \item Ramp delay~\cite{meng2010pareto} 
            \item Vehicle delay~\cite{li2016efficiency,amini2016new,kotsialos2001efficiency,kesten2013analysis,lu2017coordinated} 
            \item Weighted delay~\cite{levinson2004measuring} 
        \end{itemize} & 
        \begin{itemize}
            \item Spatial equity index ~\cite{meng2010pareto} 
            \item Gini coefficient~\cite{li2016efficiency} 
            \item Nash bargaining solution~\cite{amini2016new} 
            \item Spatial variation~\cite{kotsialos2001efficiency}  
            \item Various dispersion metrics~\cite{kesten2013analysis,lu2017coordinated} 
        \end{itemize} \\
        
        & & \\
        Perimeter control & 
        \begin{itemize}
            \item Vehicle delay~\cite{moshahedi2023alpha,aboudolas2013perimeter} 
            \item Queue length~\cite{aboudolas2013perimeter} 
        \end{itemize} & 
        \begin{itemize}
            \item AlphaFair~\cite{moshahedi2023alpha} 
            \item Dispersion~\cite{aboudolas2013perimeter} 
        \end{itemize} \\
    \end{tabular}
    \caption{\textbf{Works on fair infrastructural control}}
    \label{tab:traffic_engineering_works}
\end{table*}

\subsubsection{Fair behavioral control}
In this section we review the works from traffic engineering that explicitly approach fairness in behavioral control.
Behavioral control, including instruments such as congestion pricing, express toll lanes, license plate rationing, or tradeable mobility permits intersect a lot with the field of transportation planning, and therefore a clear distinction is renounced here.

Fairness-related works on fair behavioral control agree that equity concerns and related public acceptance issues are the Achilles Heel of behavioral control, and emphasize the need for more fairness-related studies~\cite{li2021urban,luo2022ex,yu2023income,taylor2010addressing}.
Rather than analyzing the impact on factual fairness, many works actually analyse perceived fairness through survey studies~\cite{dogterom2018acceptability,luo2022ex,finkleman2011empirical,eliasson2016congestion}.
Most studies discuss whether economic instruments in the context of behavioral control are fair at all from a qualitative perspective, and avoid discussing the quantitative question of a fairness-maximizing price at all.

In the context of tradeable mobility permits, \cite{fan2013tradable} reviews the inequities between different groups (gender, age, district, income) and takes inspiration from transportation justice ideology.
\cite{bogenberger2021mobilitycoins} studies the fairness of tradeable mobility permits in the context transportation economics \& planning aspects such as public infrastructure financing.

In the context of rationing, fairness is analysed by the dispersion of travel times~\cite{chen2020optimization,li2023booking}. 
What's more, previous works document how control gets circumvented by wealthy individuals and black markets which is problematic from a procedural fairness point of view~\cite{li2023booking,daljord2021black}.
In an extensive literature review, \cite{yu2023income} shows that almost none of the prior works on rationing analyse fairness in a quantitative way.
\cite{wang2017distributional} presents a survey on the perceived fairness and contrasts lottery- and auction-based rationing systems.

In the context of express toll lanes, three forms of fairness are distinguished: income fairness, geographic fairness and modal fairness, where income fairness is the aspect of highest attention in public debates~\cite{weinstein2004assessing,coyle2019land}. 
What's more, a differentiation between horizontal equity (fairness within groups of similar individuals) and vertical equity (fairness between different groups), as well as a classification into market, opportunity and outcome equity is made~\cite{taylor2010addressing}.
Road pricing in general is considered as regressive taxation, which means it burdens the poor relatively more than the rich~\cite{abulibdeh2018implementing}. 
\cite{schweitzer2010just} compares road pricing with other forms of regressive taxation finding that taxes that are hard to subjectively perceive, such as sales tax, cause less resistance in the public.
\cite{karlstrom2009behavioral,zhicai2008equity} study fairness quantitatively by assessing the dispersion of costs across the population.
\cite{eliasson2006equity} approaches fairness by analysing the distribution of welfare.

In addition to what was found in the fairness-related express toll lanes literature, in the context of congestion pricing, studies discuss the potential remedies of price differentiation~\cite{karlstrom2009behavioral,maheshwari2024congestion}.
In addition to a perspective on vehicles or users only, \cite{ecola2009equity} considers additional stakeholders such as residents and retail shops into the fairness debate.
Instead of asking whether congestion pricing is fair, \cite{manville2018would,manville2019longer} ask whether free (unpriced) roads can be considered fair instead, and refer to the pricing of other infrastructural utilities such as water, electricity, gas or even food, where pricing is not considered as unfair.

To summarize, most fairness-related works in the field of behavioral control are choosing a qualitative rather than quantitative way, and the few quantitative works are limited to a Egalitarian discussion related to the dispersion.
A connection to modern fairness theories is clearly missing.

\section{A Framework for Fair Traffic Engineering}
\label{sec:framework}

\begin{figure*} [!htbp!]
    \centering
    \includegraphics[width=0.8\linewidth]{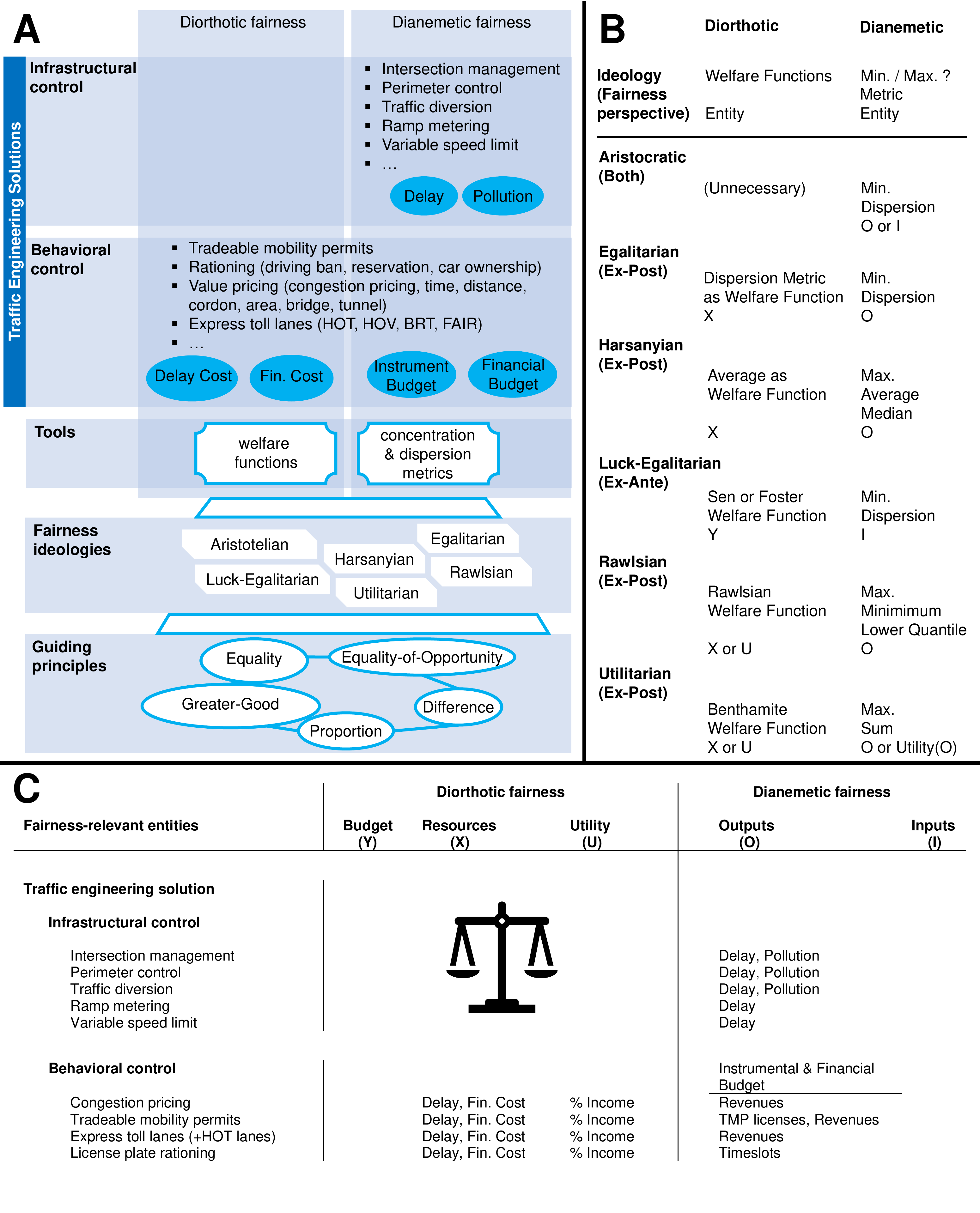}
    \caption{\textbf{Framework: Distributive Fairness for Traffic Engineering.} (A) Traffic engineering solutions and fairness-relevant resources in the context of diorthotic and dianemetic fairness, based on tools from statistics and welfare economics to assess fairness based on ideologies and guiding principles. (B) Fairness-quantifying metrics of resource allocation problems for distributive ideologies. Diorthotic (x=resources, u=utility, y=income) and Dianemetic (O=Outputs, I=Inputs) metrics for relevant ideologies. (C) A projection of the fairness quantification to the problems of traffic engineering. This table outlines fairness-relevant, teleologic entities for different infrastructural and behavioral control solutions.}
    \label{fig:riehl_quant_theory}
\end{figure*}

The distributive framework aims to enable quantitative discussions on fairness in traffic engineering by a unique connection between philosophical fairness elaborations and algorithmic traffic engineering solutions.
After defining purpose \& properties, we demarcate the framework from the field of transportation justice.
Next, we identify relevant philosophical ideologies \& moral guiding principles for the discussion at hand, and propose tools to quantify fairness.
Then, we outline how traffic engineering solutions can be modelled as resource allocation mechanisms, determine fairness-relevant resources, and connect these insights with the ideologies. 
Fig.~\ref{fig:riehl_quant_theory} summarizes the structure of the proposed framework.

\subsection{Purpose \& Properties}

The proposed fairness framework is developed for the domain of traffic engineering and shall serve two purposes:
(i) it shall be useful when designing traffic engineering solutions in terms of fairness without sacrificing transportation efficiency; 
(ii) it shall be convincing, and equip traffic engineers with a urgently-needed, useful, argumentative tool-set to confront equity considerations in public debates.

To serve these purposes, the framework must discuss fairness in a quantitative, ideology-free, distributive, and pragmatic way.
A quantitative discussion of fairness enables the integration of fairness in the optimization problems of the automated, algorithm-driven approach of traffic engineering. A conceptual, transcendental discussion of fairness is insufficient.
In order to be convincing, the framework must be ideology-free and allow for the integration of various different fairness ideologies, rather than advocating a specific one. This allows for a holistic and systematic analysis of fairness from different conceptual perspectives.
For a quantitative discussion, we choose to model traffic engineering solutions as resource allocation mechanisms, and focus on teleologic, distributive fairness ideologies.
In order to be of value, efforts to improve fairness must be pragmatic, and cannot come at the cost of transportation efficiency or cause even more congestion than there already is.

\subsection{Demarcation against transportation justice}
The framework addresses traffic engineering that is fundamentally different from the commonly discussed fairness aspects in transportation justice, which relates to transportation planning, accessibility and potential mobility.
While conventional transportation planning disregarded the improvement of public transportation systems in favor of investments into individual mobility in the form of road transportation systems in many parts of the world, traffic engineering is limited to controlling existing road transportation systems only. 
From this point of view, our framework is of little use to contribute to any questions related to planning and public transportation expansion.
However, it can attribute public transportation with dedicated lanes or privileged treatment in control-algorithms to achieve prioritization.
The authors and the proposed framework fully agree with the valid concern of the transportation justice movement: \textit{"accessibility shortfalls caused by congestion are as much an injustice as accessibility shortfalls caused by inadequate public transport services or any other cause"}(Karel Martens~\cite{martens2012justice}).


\subsection{Relevant fairness ideologies}
Many different ideologies evolved in the history of ethics \& philosophy; disparities and fairness do not necessarily exclude each other and are rather advocated to a certain extent. 
While all ideologies have their argument, they differ in their specific interpretation of what fairness means~\cite{goppel2016handbuch}.
We identified six ideologies and five guiding principles on fairness which are particularly useful for this distributive framework, and propose their use for quantification: Aristotelian (proportion principle, also "Aristocratic"), Utilitarian (greater-good principle), Rawlsian \& Harsanyian (difference principle), Egalitarian (equality principle), and Luck-Egalitarian (equality-of-opportunity principle) ideology.
The proportion principle describes an allocation as fair, as long as it is proportional to the status of individuals. 
The status could reflect the individual's contributions (e.g. tax paid, income, economic power), or could reflect the individual's need (e.g. urgency).
The greater good principle describes an allocation as fair, as long as it maximizes the greater good of the society. 
This can also imply the sacrifice of the few for the benefit of the many. 
The difference principle describes an allocation as fair, as long as it achieves the best possible outcome for the least-advantaged (Rawlsian) resp. the best possible average outcome (Harsanyian).
The equality principle describes an allocation as fair, as long as it achieves equality of endowment.
The equality-of-opportunity principle describes an allocation as fair, as long as the initial chances / opportunities of each individual as inputs to a (stochastic) societal process have been equal.
For sure, the application of these ideologies must be discussed in the context of a specific resource allocation problem, and not all of these ideologies might equally-well apply in every context. 

\subsection{Quantifying dianemetic and diorthotic fairness}
In order to translate the qualitative ideologies to quantitative metrics, we propose the use of Welfare functions for diorthotic, and concentration \& dispersion metrics for dianemetic fairness elaborations.
In terms of dianemetic fairness, ideologies differ in their perspective on the relevance of individual outputs $O$ and inputs $I$ to a resource allocation process. 
The inputs $I$ can refer to contributions or social status (Aristocratic), or to starting conditions (Luck-Egalitarian).
The outputs $O$ can refer either to allocated resources, or the actual utility the resources have to the individuals.
In terms of diorthotic fairness, a discussion on market equilibria, and Welfare economics is proposed. 
Market equilibria emerge from a resource allocation process in a population of individuals with a free will that aim to achieve the best possible outcome for themselves (rationality assumption).
In this context, each individual possesses a budget $y$, to acquire resources $x$ via transactions, that generate a certain utility $u$ for the individual, depending on its personal preferences and utility functions.
One can stress Market equilibria towards Welfare-maximizing optima, for example by introducing (redistributive) taxation or pricing. 
These optima are located at the tangential point of bliss between Pareto-efficient, market-equilibria frontiers and Welfare functions. 
Welfare functions define the understanding of what Welfare is, and can be tailored to reflect fairness considerations, and depending on the ideology include allocated resources $x$, budgets $y$ or utility $u$.
Further details on how to formalize Welfare functions, and concentration \& dispersion metrics can be found in the Supplementary Note 3.

\subsection{Modelling as resource allocation mechanism}

Traffic engineering can be considered as the allocation of spatio-temporal resources.
Infrastructural control can be considered as institution in the context of dianemetic fairness.
Behavioral control has both components: the diorthotic component to assess whether the (market) interactions between users and resulting market equilibria are fair, and the dianemetic component to assess whether the initial endowment can be considered fair (a prerequisite for fair, diorthotic transactions). 
In order to discuss distributive fairness in the context of traffic engineering, fairness-relevant resources \& actors must be identified.
With regards to philosophic theory, we find which criteria can be used to determine the relevance to fairness.
Rawls argues that primary goods, that are of interest and affect utility for every human - independent of their preferences - are relevant to fairness~\cite{rawls1971atheory}.
Walzer argues that goods with a distinct social meaning need to be distributed in dedicated, fair, distributive spheres, contrary to normal goods~\cite{walzer1983}. 
The capability approach advocates to discuss resources that determine capabilities (freedom of choice through many opportunities) rather than functionings (actual outcomes)~\cite{nussbaum2011creating,sen2008idea}.

In the context of infrastructural control, we consider delays and pollution as fairness-relevant resources. 
Delay can be considered as the difference between the free-flow travel time and the actual travel time a user experiences. 
Plain delays, perceived delays (over-proportionally increasing with longer durations), as well as delays per travelled distance can be subject to discussion. 
Delays are undesired by every rational passenger, cause costs to them (negative utility depending on urgency), can accumulate to a significant amount of life-time and therefore lost opportunities. 
Traffic related externalities (pollution) causes damages to health \& environment (noise, emissions, safety), and affects locations (and people living or working there) rather than vehicles.

In the context of behavioral control, we consider financial \& delay costs (diorthotic), and financial \& instrumental budgets (dianemetic) as fairness-relevant resources.
Behavioral control instruments incentivise a change in the choice behavior of users (modality, route, timing, and temporal-scope) in order to overcome selfish preferences and to achieve more desirable decisions (on system-level).
The modality choice refers to the means of travel (e.g. single car, carpool, public transport, staying at home).
The route choice refers to the route of travel (e.g. tolled but fast highway vs. taking a free but slower route).
The timing choice refers to when certain travels are conducted (e.g. on-peak vs. off-peak-hours), and the choice of starting times.
The temporal-scope choice refers to the trade-off between present and future delay costs.
Behavioral control solutions achieve a change in choice behavior by introducing financial costs to the users, allowing them to trade-off financial and delay costs on a market-like process from a diorthotic point of view.
Budgets can refer to financial means, but also can they refer to instrumental budgets, such as ration cards \& permits, redistributed, monetary revenues, or artificial currencies.
Financial means are usually determined by labor work opportunities; both can usually be seen as fixed on the short-term perspective, and are observed to be unequally distributed across most modern populations. 
Due to the temporal horizon of traffic engineering, one could therefore argue that dianemetic discussions are rather related to transportation planning than to traffic engineering.
However, deftly design of behavioral control instruments and redistributive measures can address these equity issues as well.

We advocate context-, resource- and investigation-specific scoping when defining fairness-relevant actors.
In general, we argue that the ultimate goal must be a fair distribution of relevant resources across single users. 
In practice however, an aggregated view on vehicle or vehicle groups might be more feasible and pragmatic.
For example, this could involve a discussion on vehicles of the same type (e.g. public transport vehicles, bicycles, delivery vehicles), or vehicles of the same origin (e.g. vehicles from a certain suburb or region that go to the city), or vehicles of the same purpose (e.g. work commute, recreational).
Moreover, depending on the resource discussed, different relevant actors or actor groups must be identified. 
For cost resources with negative utility, namely externalities of road transportation such as pollution, users that experience the costs should be considered. 
For pollution, it is less of the users of a road transportation network, and rather specific locations or local resident groups that need to be taken into account.
Moreover, further external stakeholders such as residents, or retail shops can be taken into account.



\subsection{Projection of fairness to traffic engineering}

We made two important observations, when bringing traffic engineering solutions and quantitative measures for dianemetic and diorthotic fairness together. 
First, we do not see any fairness-relevant resources that could be considered as an input ($I$ and $y_i$) to a resource allocation process and could be addressed by existing traffic engineering solutions. 
However, in future systems with autonomous and fully-connected vehicles, inputs such as bids for auction controllers could be possible.
Thus, only teleological (ex-post) fairness theories apply, and Aristocratic and Luck-Egalitarian ideologies are excluded from this discussion.
Second, a discussion on delay, pollution \& financial cost resources deviates from usual fairness discussions in that they are of negative utility to the users. 

While delay \& pollution resources could be discussed for any infrastructural control, we argue that pollution is usually a concern in control applications in an urban context, meaning intersection management, perimeter control and traffic diversion only.
Delays can be translated to monetary values using subjective value of time (VOT) which allows the trade-off with financial costs.
These cost resources could be translated to utilities in order to discuss costs as standardized monetary values of the income (share of income).

Instrument-specific budget resources could refer to entities such as licenses in the context of tradeable mobility permits, and timeslots in the context of license plate rationing. 
What's more, generated revenues (from congestion pricing, express toll lanes, tradeable mobility permits) could be redistributed across the population and would thus affect the financial budgets.
These generated revenues could also be invested into public transportation and would thus affect available alternatives.

\section{Demonstration Case Studies}
\label{sec:casestudies}

Two case studies for infrastructural \& behavioral control outline how fairness can be taken into consideration when designing traffic engineering solutions, by applying the proposed framework. Moreover, the case studies serve to assess the relationship between fairness and efficiency, and to compare the various different measures used to quantify fairness from different ideologies perspective.

\subsection{Key Insights}
The results show, that fairness \& efficiency do not categorically exclude each other, and that higher levels of fairness can be achieved without sacrificing transportation efficiency.
It can be observed that Utilitarian, Rawlsian, and Harsanyian ideology expose a similarity with transportation efficiency, while Egalitarian ideology contradicts the most.
A dendrogram analysis of different fairness and efficiency measures reveals meaningful, consistent clusters in the proposed, fairness-quantifying measures, that correspond to theoretical fairness ideologies across traffic engineering applications.

\subsection{Signalized intersection management}

\begin{figure*} [!htbp!]
    \centering
    \includegraphics[width=0.8\linewidth]{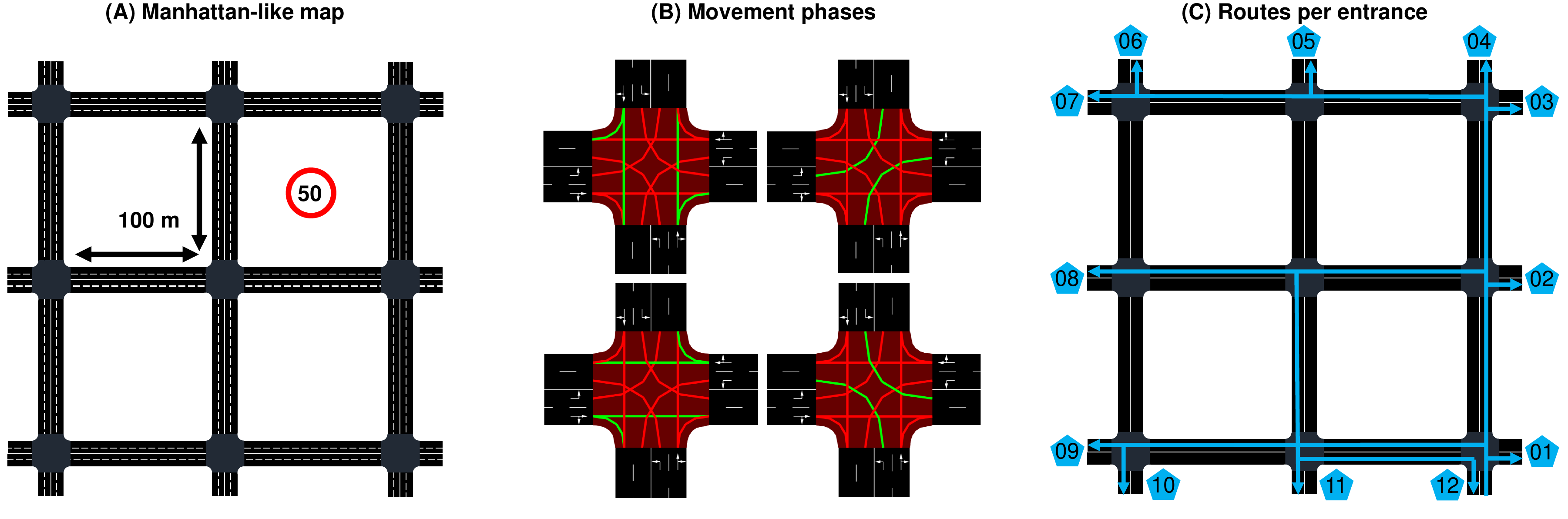}
    \includegraphics[width=0.8\linewidth]{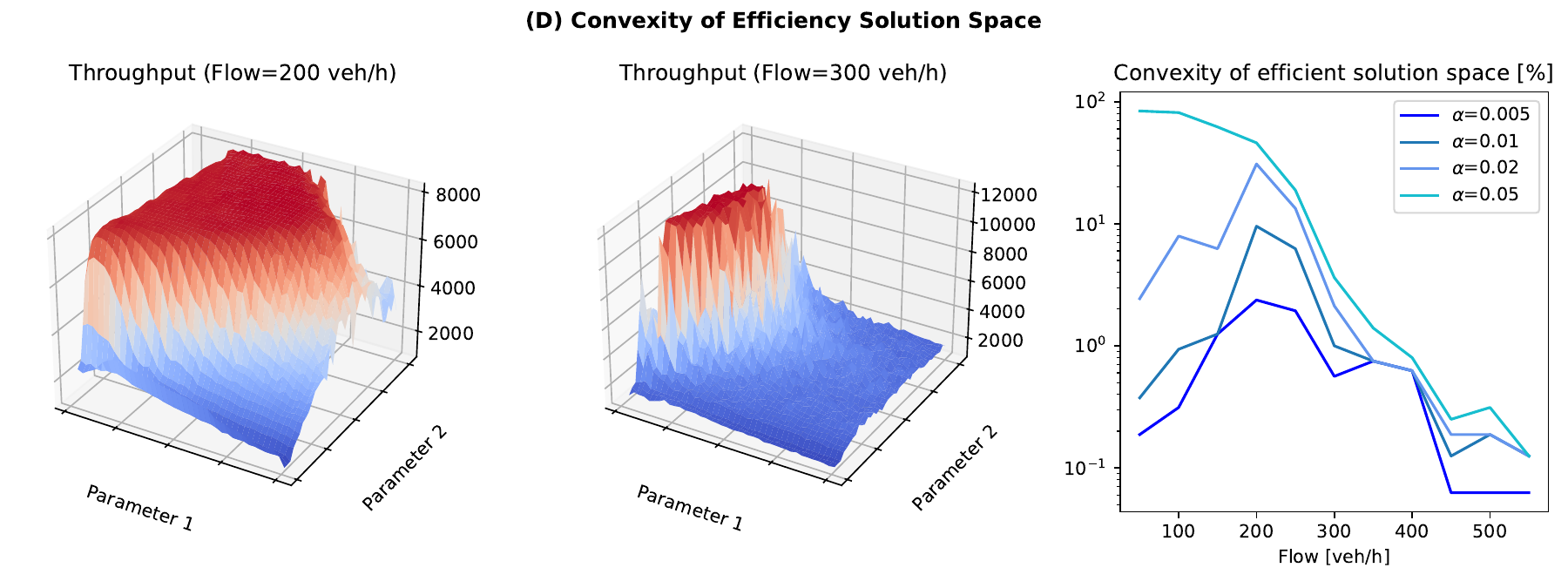}
    \includegraphics[width=0.8\linewidth]{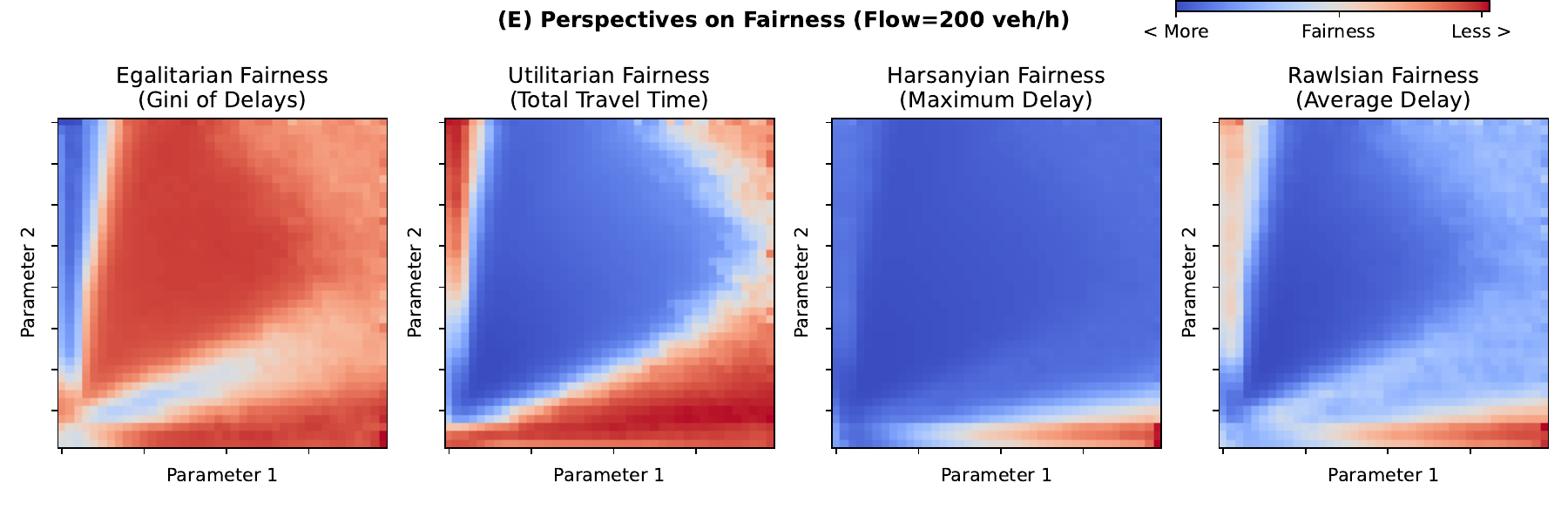}
    \includegraphics[width=0.4\linewidth]{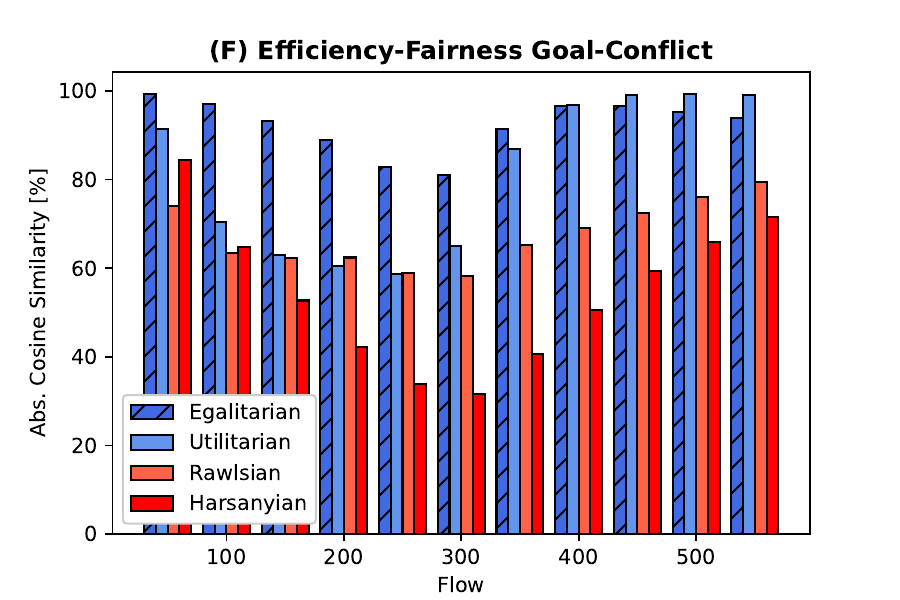}
    \includegraphics[width=0.4\linewidth]{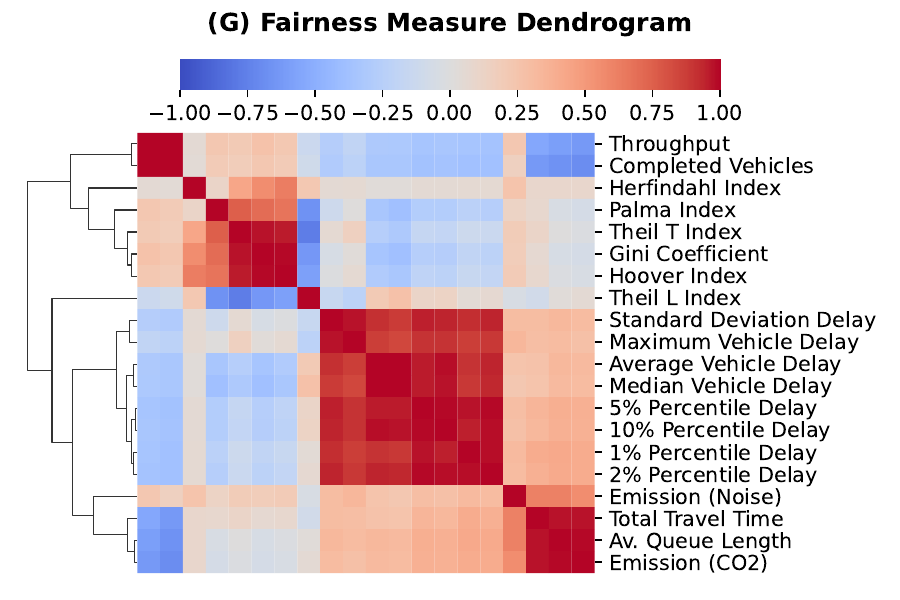}
    \caption{\textbf{Case Study: Signalized Intersection Management}. \\
    (A) The Manhattan-like road network for the case study consists of nine intersections, that are connected with four-laned, bi-directional, 100m long roads. (B) Each intersection is equipped with a fixed-cycle traffic light and four distinct movement phases. The transition from one to the next movement phase takes 3 seconds of transient yellow time for safety reasons. (C) The road network has twelve entrances \& exits. An equal, static traffic flow at each of entrances is generated, and spawned vehicles are randomly assigned one of the twelve possible routes with an even probability distribution. (D) Two design parameters for the traffic light controller span the solution space: the green durations of the straight and turning movement phase. Each parameter is varied from one to 40 seconds. Equal parameters for each intersection are assumed, due to the network symmetry. In order to enable a more fluent traffic, time-shifts are applied between neighboring cycles; the time-shift equals the average of the parameters. The $\alpha$-efficient space is a subset of the solution space, in which the efficiency equals at least $(1-\alpha)$ of the maximum efficiency possible. The convexity is measured as the ratio of powers of $\alpha$-efficient and solution space. The smaller the flows, the higher the room for improvement of fairness. (E) Four teleological fairness ideologies measured across the solution space. (F) The goal-conflict between fairness and efficiency is measured by the Cosine Similarity for different traffic flows (larger Cosine Similarity means less goal conflicts). Hatched bars (Egalitarian) are negative. (G) A metric dendrogram of various recorded efficiency and fairness metrics. The clusters show that the many fairness measures form clusters that correspond to the four underlying fairness theories.}
    \label{fig:casestudy_trafficlight}
\end{figure*}

The first case study shows how dianemetic fairness can be taken into account when designing infrastructural control, namely signalized intersection management as outlined in Fig.~\ref{fig:casestudy_trafficlight}.
The static design problem is concerned with an uncoordinated, fixed-cycle traffic light controller with two integer parameters.
A common measure for the transportation efficiency of intersections is the number of passing vehicles per time (throughput).
The setup can be considered as dianemetic, institutional endowment resp. allocation of delay resources by the traffic light.

How can we improve the fairness of the signalized intersection management?
When optimizing for transportation efficiency, the $\alpha$-efficient solution space (efficiency equals at least $1-\alpha$ of the maximum efficiency) leaves room for fairness considerations, and in cases of lower network saturation allows for significant improvements in fairness without sacrificing efficiency.
This of course depends on whether fairness and efficiency form a goal conflict, and how large the  $\alpha$-efficient solution space resp. how convex the transportation efficiency is. 
Our results shows, that the Egalitarian fairness ideology is conflicting strongly with different measures of efficiency, while the other fairness ideologies prove to have a weak conflict (depending on the network flow).
Moreover, the convexity of the transportation efficiency is observed to decrease for larger flows in the network, hence fairness improvements can especially achieved at lower levels of saturation.

\subsection{Static Road Pricing}

\begin{figure*} [!htbp!]
    \centering
    \includegraphics[width=0.4\linewidth]{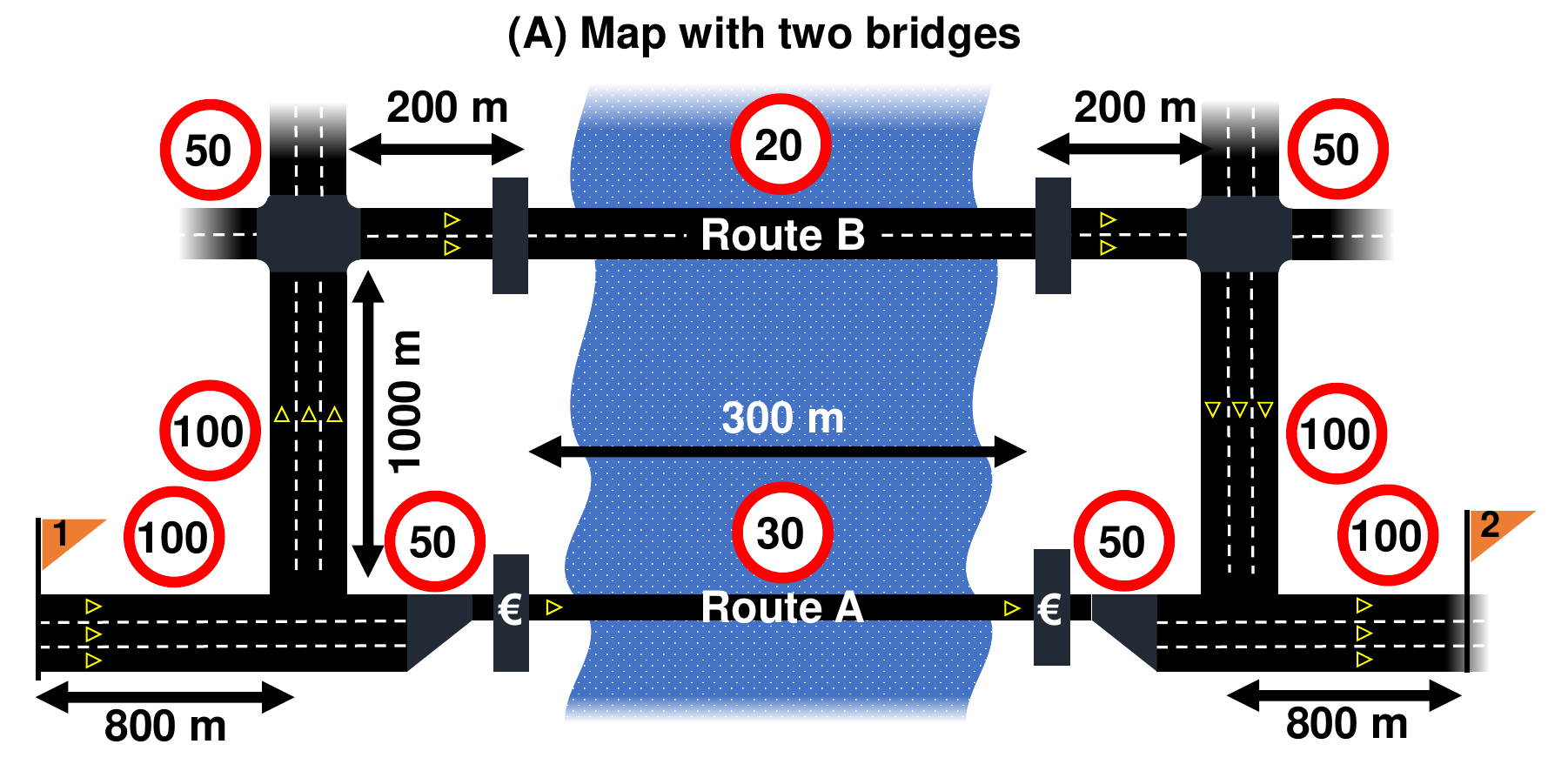}
    \includegraphics[width=0.4\linewidth]{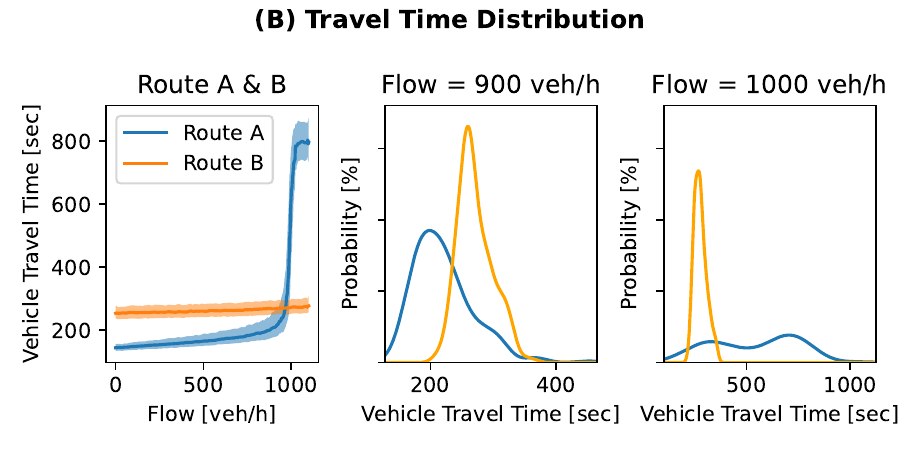}
    \includegraphics[width=0.4\linewidth]{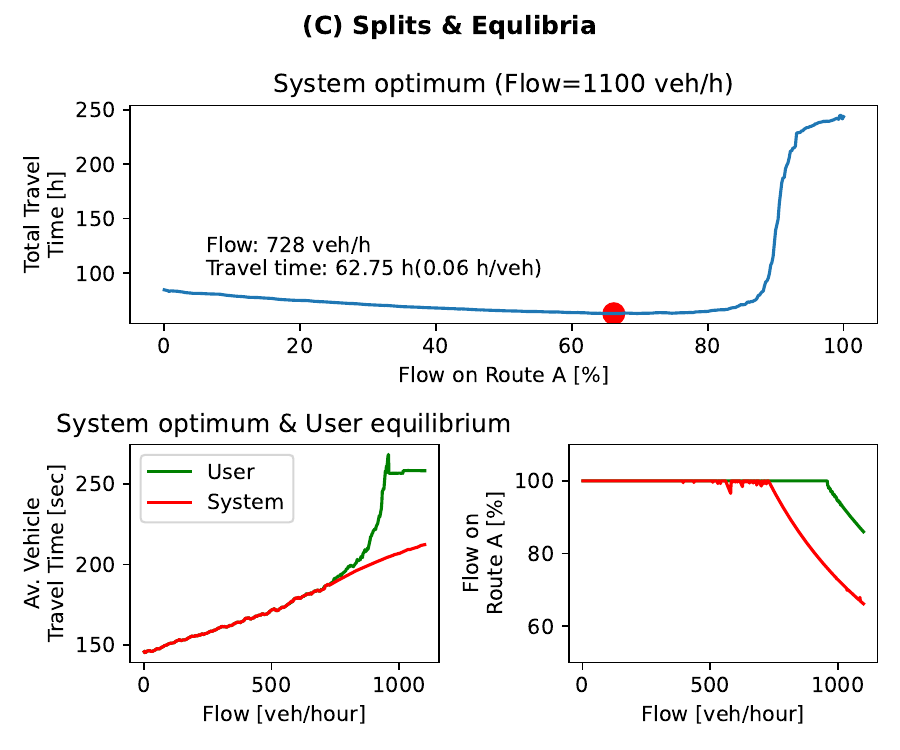}
    \includegraphics[width=0.4\linewidth]{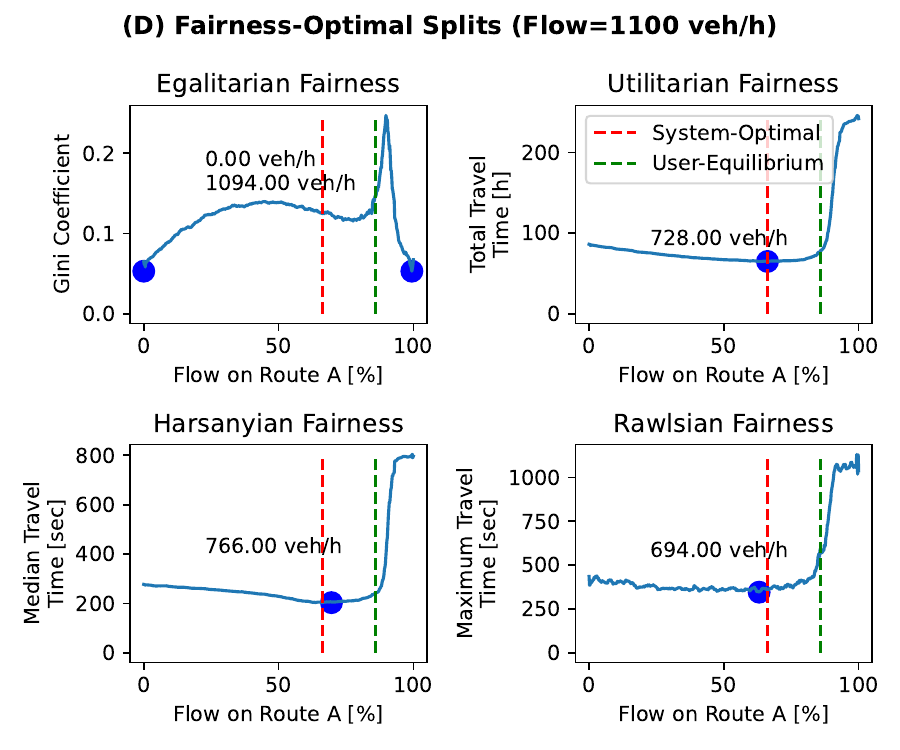}
    \includegraphics[width=0.8\linewidth]{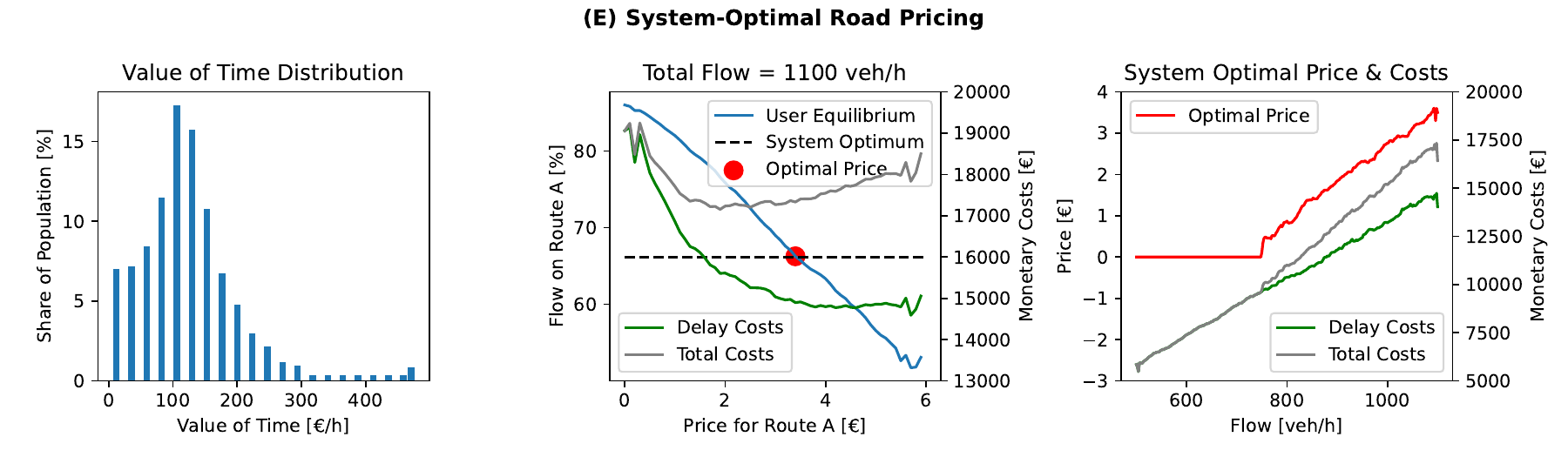}
    \includegraphics[width=0.4\linewidth]{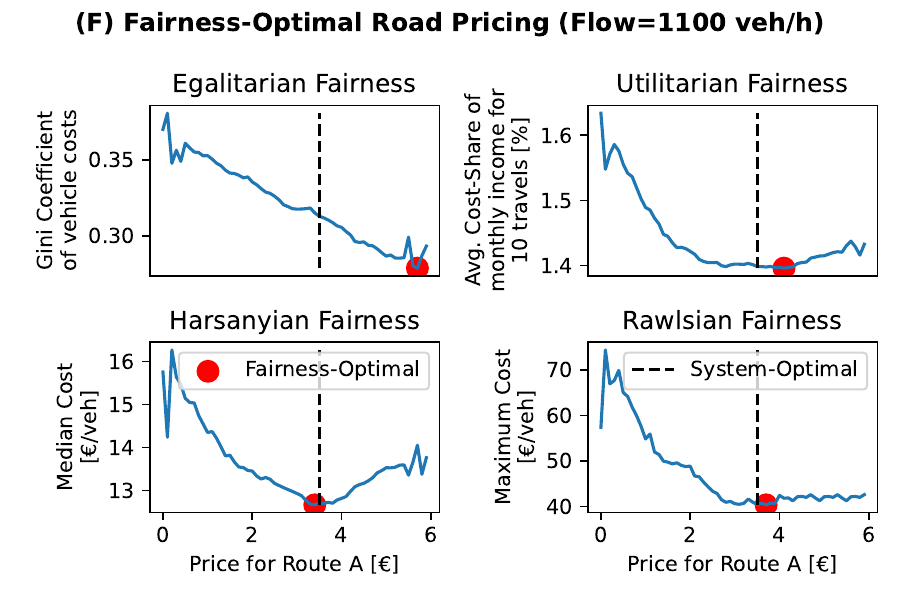}
    \includegraphics[width=0.4\linewidth]{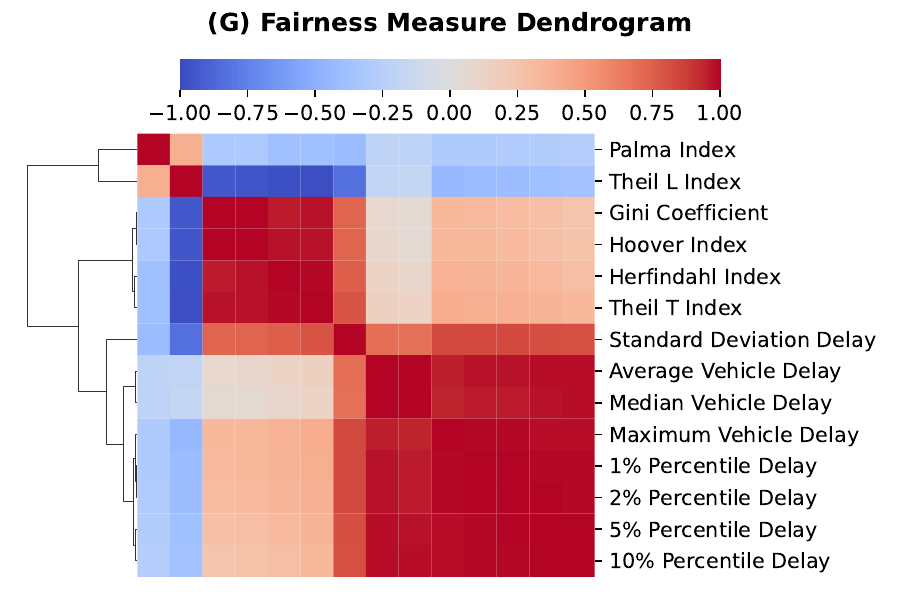}
    \caption{\textbf{Case Study: Static Road Pricing}.\\
    (A) The road network for this case study consists of a starting point 1 and destination point 2 that are separated by a river. Two possible routes A \& B (bridges or tunnels) are available. Route A is short, but has the bottleneck of a single-lane. Route B is longer, as it requires to drive one additional kilometer on each riverbank but is two-laned. (B) The travel time distribution for each of the two routes at different flows. Route A offers a faster alternative to B, but reacts sensitive to higher flows and reaches congested regimes after around 950 vehicles per hour. Route B is slower in general, but offers a travel time that is more robust to increasing flows. The travel times are not-normally distributed and right skewed in congested regimes. (C) System-optimal splits and Wardrop-equilibrium (user-optimum) splits. Being left uncontrolled, the route choice behavior of a population of rational users causes suboptimal, higher total and average travel times. (D) Fairness-optimal splits determined as tangential (point of bliss) between fairness-related, social indifference curves (Welfare functions) and the Pareto-efficient front (abscissa). Please note, fairness here is in terms of the distribution of delays (travel times) and not monetary values. (E)  By surcharging a fee for taking route A, users are increasingly incentivised to choose route B. We assume that users are heterogeneous in their readiness to pay for route A, according to their value of time (VOT). System-optimal road pricing transitions the Wardrop-equilibrium to a system-optimum. (F) Fairness-optimal road pricing. Pricing improves fairness across all ideologies. Except for the Egalitarian fairness ideology, all fairness ideologies have a fairness-optimal price that is close to the efficiency-optimal price. Egalitarian ideology however, is the most common measure for fairness in current literature, which is highly problematic. (G) A metric dendrogram of various recorded efficiency and fairness metrics. The clusters show that the many fairness measures form clusters that correspond to the four underlying fairness theories.}
    \label{fig:casestudy_congestionpricing}
\end{figure*}

The second case study is concerned with diorthotic fairness and the design of behavioral control, namely static road pricing as outlined in Fig.~\ref{fig:casestudy_congestionpricing}.
In this case study, the pricing of a bridge is used to control the route choice behavior (split) of the population across two available route options.
The Wardrop equilibrium~\cite{wardrop1952road} is a mechanism that describes how splits take place in absence of any price as the result of market-like interactions between individuals of a population. 
Rational users are observed to act selfish, and to choose routes that minimize their own travel time. 
As a consequence, a population of rational users is observed to form a split where both routes have an equal travel time, and there is no opportunity to improve one's situation by changing.
From a system point of view, the split can be considered optimal when the total travel time (measure for efficiency) is minimized. 
For flows larger than 750 vehicles per hour, the selfish behavior of drivers leads to splits that are sub-optimal from a system perspective, and unnecessarily increases both total and average travel time.

How fair are splits at system optimum and Wardrop-equilibrium?
From an Harsanyian, Rawlsian and Utilitarian perspective, splits around the system-optimum can be considered as fairness-maximizing. 
The Egalitarian ideology suggests two possible splits; either all vehicles shall be allocated to route A or all vehicles shall be allocated to route B, as this leads to a minimal dispersion in travel times, as all drivers would experience most similar travel times. 
This suggestion seems quite radical, as it completely overlooks the benefits of splitting the flow.
Taken together, all fairness ideologies consider the Wardrop-equilibrium as unfair.
By introducing financial costs, users are enabled to trade-off financial and delay costs in a way, that the user equilibrium moves towards the system optimum.
For a price of around 3.50 \texteuro the Wardrop-equilibrium reaches the system-optimum, and fairness-maximizing delay distribution is achieved.

However, the users now experience two types of costs: (i) costs due to the surcharge fee (for route A users only), and (ii) costs due to the delays (product of VOT and delay times).
The additional financial burden to the user raises the question whether controlling the behavior of the population towards the system-optimum split is fair to the users, and whether it generates value for those.
The introduction of such a surcharge can significantly reduce the total costs and delay costs, as the pricing-incentive leads to more travel-time-optimal flow allocations.
However, from a price of 2.00 \texteuro onwards, the total costs (including financial fee costs and the travel time costs) increase again. 
While from a transportation efficiency perspective it makes sense to go even further to higher prices, from a financial perspective the benefits of congestion pricing decrease after this price.
While congestion pricing can decrease travel time costs, it increases financial costs arbitrarily and hence the total costs.
The financial costs, and the gap between total costs and delay costs, increases with larger flows.

So which price achieves a fair distribution of total costs rather than just delays? 
Compared with the situation of no congestion pricing (0.00 \texteuro), the system-optimal pricing leads to a higher fairness in terms of all four fairness ideologies.
However, the fairness ideologies suggest slight deviations from this system optimal price to achieve fairness optimality.
Again, the Egalitarian fairness ideology aims to avoid a split and to direct the whole flow to only one route.

\section{Conclusions}
\label{sec:conclusions}

This work highlighted the importance of fairness in road traffic engineering.
The review of previous work on fair traffic engineering revealed that there is lack of connection with the fairness literature, unsystematic discussion of fairness, and the employment of over-simplifying definitions of fairness.
As a result, the necessity for a framework to quantitatively assess fairness was identified.
The study then proposed a distributive, quantitative, ideology-free, and pragmatic fairness framework for infrastructural and behavioral traffic control systems, which was showcased and demonstrated at the example of two case studies.

Future work is encouraged to exploit this useful, quantitative framework when designing infrastructural and behavioral traffic engineering solutions and to address the need for an integrative, multi-perspective view on fairness. 
More extensive studies on the relationship of fairness and efficiency could quantitatively address different fairness ideologies. 
Studies on behavioral control could not only assess whether economic instruments are fair, but also determine fairness-optimizing prices, and explore fairness-optimizing price differentiation. 
Last but not least, the exploration of perceived fairness rather than objective fairness (e.g. perceived delays vs factual delays) is a promising research direction.


\subsection*{CRediT authorship contribution statement}
\textbf{Kevin Riehl: } Conceptualization, Methodology, Formal analysis, Investigation, Data Curation, Writing - Original Draft, Visualization, Project administration. 
\textbf{Anastasios Kouvelas: } Writing - Review \& Editing, Validation.
\textbf{Michail Makridis: } Writing - Review \& Editing, Validation, Supervision.

\subsection*{Declaration of competing interest}
None.

\clearpage
\renewcommand*{\bibfont}{\normalfont\small}
\printbibliography

\end{document}